\documentclass[reprint,amsmath,amssymb, aps]{revtex4-2}
\usepackage{multirow}
\usepackage{amssymb}
\usepackage{amsmath, graphicx}
\usepackage{dcolumn}
\usepackage{bm}
\usepackage{enumerate}
\usepackage{slashed}
\usepackage{epstopdf}
\usepackage[unicode]{hyperref}
\usepackage{csquotes}
\usepackage[usenames]{color}

\newcommand{\be}{\begin{equation}}
\newcommand{\ee}{\end{equation}}
\newcommand{\bea}{\begin{eqnarray}}
\newcommand{\eea}{\end{eqnarray}}
\newcommand{\ben}{\begin{enumerate}}
\newcommand{\een}{\end{enumerate}}
\newcommand{\bde}{\begin{widetext}}
\newcommand{\ede}{\end{widetext}}

\newcommand{\Tr}{\mathrm{Tr}}

\newcommand{\bc}{\begin{center}}
\newcommand{\ec}{\end{center}}

\usepackage{braket}

\begin{document}

\title{\boldmath Insights of quantum time into quantum evolution}

\author{Ngo Phuc Duc Loc}
\email{locngo148@gmail.com}

\affiliation{Department of Physics and Astronomy, University of New Mexico,  Albuquerque, NM 87131, USA}

\begin{abstract}
If time is emergent, quantum system is entangled with quantum time as it evolves. If the system contains entanglement within itself, which we can call internal entanglement to distinguish it from the ``external" time-system entanglement, the speed of evolution is enhanced. In this paper, we show the correlation between the novel time-system entanglement and the conventional internal entanglement of a system that contains two entangled qubits. We consider two cases: (1) two initially entangled qubits that evolve under local dynamics; (2) two interacting qubits such that entanglement between them is generated over time. In the first case, we obtain the main result that increasing internal entanglement speeds up the evolution and makes the system more entangled with time. For both cases, we show the dependence of time-system entanglement entropy on the distance of evolution which is characterized by fidelity. The interacting system can evolve faster than the non-interacting system if the interaction is sufficiently strong, and thus it can be entangled with time more efficiently.
\end{abstract}
\maketitle

\section{Introduction}
The nature of time is a big puzzle. Does time have a quantum structure? Or, in other words, is it emergent? The physics community recently is interested in ``extracting space" from entanglement \cite{sean1,sean2,Giddings}, but what about time? Given the fact that space and time are intimately connected in relativity theory, it is therefore not unreasonable to ask for the emergence of time, just like the emergence of space. Unlike space, however, the arrow of time must somehow emerge as well.

Entanglement is another mysterious feature of the quantum world. While the nature of entanglement is still an open question at the fundamental level \cite{aspect}, it is both interesting and important to understand various phenomenological aspects of entanglement. Entanglement could be generated by interaction between systems, or by interaction between a system and an environment that is crucial to understand decoherence \cite{joos}. Entanglement can also be ``artificially felt" by different observers when a quantum state is Wigner rotated due to a Lorentz transformation \cite{Wigner1,Wigner2}.

If time really has a quantum structure, another kind of entanglement arises: quantum system is entangled with quantum time as it evolves. For the classic works, see \cite{PaW,gio,marletto}. There are some studies of quantum time in special relativity \cite{loc,singh}, in quantum field theory \cite{diaz1,diaz2,diaz3}, in quantum gravity \cite{carlo,zeh,sal}, in cosmology \cite{kiefer}, and in the context of decoherence \cite{sebas}. For the experimental prospects of this quantum time idea, see \cite{exp1,exp2,exp3}.

In this so-called Page-Wootters formalism (PaW), the global state of the Universe, $\ket{\Psi}\rangle$, is static and lives in the total Hilbert space $\mathcal{H}_{tot}$. The quantum state of the system, $\ket{\psi(t)}$, evolves as time passes and lives in the system Hilbert space $\mathcal{H}_S$. The quantum state of time, $\ket{t}$, lives in the time Hilbert space $\mathcal{H}_T$. Schematically, one has $\mathcal{H}_{tot}=\mathcal{H}_T\otimes\mathcal{H}_S$ and $\ket{\Psi}\rangle\propto\int\ket{t}\otimes\ket{\psi(t)}dt$. In the PaW formalism, the main spirit is that: the more the system evolves, the more it becomes entangled with time.

The quantum system itself may also contain entangled subsystems. Besides the relevance in quantum information protocols, this is also typical in nature. For example, a quantum state of black hole describes entanglement between interior and exterior degrees of freedom \cite{Hawking}. Another example is a quantum state that describes entanglement between subhorizon and superhorizon modes of cosmological perturbations \cite{loc2,EE}. This motivates us to study the connection between the conventional entanglement within the system, which we will call ``internal entanglement" from now on, and the novel ``external" time-system entanglement mentioned above. As we will show, this connection is possible due to the fact that increasing internal entanglement speeds up the evolution and thus enhances the time-system entanglement. As a first step to flesh out the idea, we will consider two simplified cases: (1) two initially entangled qubits that evolve under some local dynamics; (2) two interacting qubits such that their internal entanglement is generated as time passes.

This paper is organized as follows. In Sec. \ref{sec_non-interacting}, we consider the case of two initially entangled but non-interacting qubits. We first show that increasing internal entanglement can make the state vector travel further and faster.  We then compute the time-system entanglement measures when there is a single qubit clock. We show that increasing internal entanglement speeds up the evolution and thereby makes the system more entangled with time. We then generalize our results to the continuous limit. In Sec. \ref{sec_interacting}, we study time-system entanglement for a system containing two interacting qubits, focusing on computing the time-system entanglement entropy in the continuous limit. We show that the interacting system can evolve faster than the non-interacting system if the interaction is sufficiently strong, and thus it can be entangled with time more efficiently. Summary and possible further developments are presented in Sec. \ref{sec_conclusions}.

\section{Two non-interacting qubits}\label{sec_non-interacting}

\subsection{Speed of evolution}
Consider the initial quantum state of two entangled qubits \cite{27}:
\begin{equation}\label{initial state}
    \ket{\psi(0)}=\alpha\ket{00}+\beta\ket{11},
\end{equation}
where $\alpha$ and $\beta$ are complex numbers satisfying the normalization condition $|\alpha|^2+|\beta|^2=1$. The basis vectors are energy eigenstates of the local Hamiltonians:
\begin{equation}\label{local_Hamiltonian1}
    H_A\ket{0}_A=0,\hspace{1cm}H_A\ket{1}_A=\epsilon\ket{1}_A,
\end{equation}
\begin{equation}\label{local_Hamiltonian2}
    H_B\ket{0}_B=0,\hspace{1cm}H_B\ket{1}_B=\epsilon\ket{1}_B,
\end{equation}
where the subscripts indicate the corresponding subsystems and $\epsilon>0$. The total Hamiltonian is given by
\begin{equation}
    H_{total}=H_A\otimes I_B+I_A\otimes H_B.
\end{equation}
This Hamiltonian acts locally on each subsystem and therefore entanglement measures are preserved throughout the state's evolution.

The entanglement entropy of the quantum state in Eq. \ref{initial state} is
\begin{equation}\label{S(A)}
\begin{aligned}
    S(A)&=-\Tr(\rho_A\log_2\rho_A)\\
    &=-\left(|\alpha|^2\log_2|\alpha|^2+(1-|\alpha|^2)\log_2(1-|\alpha|^2)\right).
\end{aligned}
\end{equation}
Another useful entanglement measure is the quadratic entanglement entropy which is determined from the purity of the state instead of its eigenvalues. This quantity may be more easily accessible experimentally \cite{tanaka}. It is given by
\begin{equation}\label{S2(A)}
    S_2(A)=2(1-\Tr(\rho_A^2))=4|\alpha|^2(1-|\alpha|^2).
\end{equation}

The state vector at time $t$ is
\begin{equation}
    \ket{\psi(t)}=\alpha\ket{00}+\beta e^{-2i\epsilon t/\hbar}\ket{11}.
\end{equation}
The fidelity (or overlap) between the initial and final states is
\begin{equation}\label{fidelity}
\begin{aligned}
    \Delta\psi&\equiv|\braket{\psi(t)|\psi(0)}|\\
    &=\sqrt{1+2|\alpha|^2(1-|\alpha|^2)\left(\cos{\frac{2\epsilon t}{\hbar}}-1\right)}.
\end{aligned}
\end{equation}
Smaller fidelity means larger distance, so it can also be thought of as a distance measure. We define $\tau$ to be the amount of time for the state to travel through the distance $\Delta\psi$. For convenience, it is given in units of $\hbar/\epsilon$:
\begin{equation}\label{tau}
    \frac{\tau}{\hbar/\epsilon}=\frac{1}{2}\arccos\left(\frac{(\Delta\psi)^2-1}{2|\alpha|^2(1-|\alpha|^2)}+1\right).
\end{equation}

\begin{figure}[h!]
    \centering
    \includegraphics[scale=0.8]{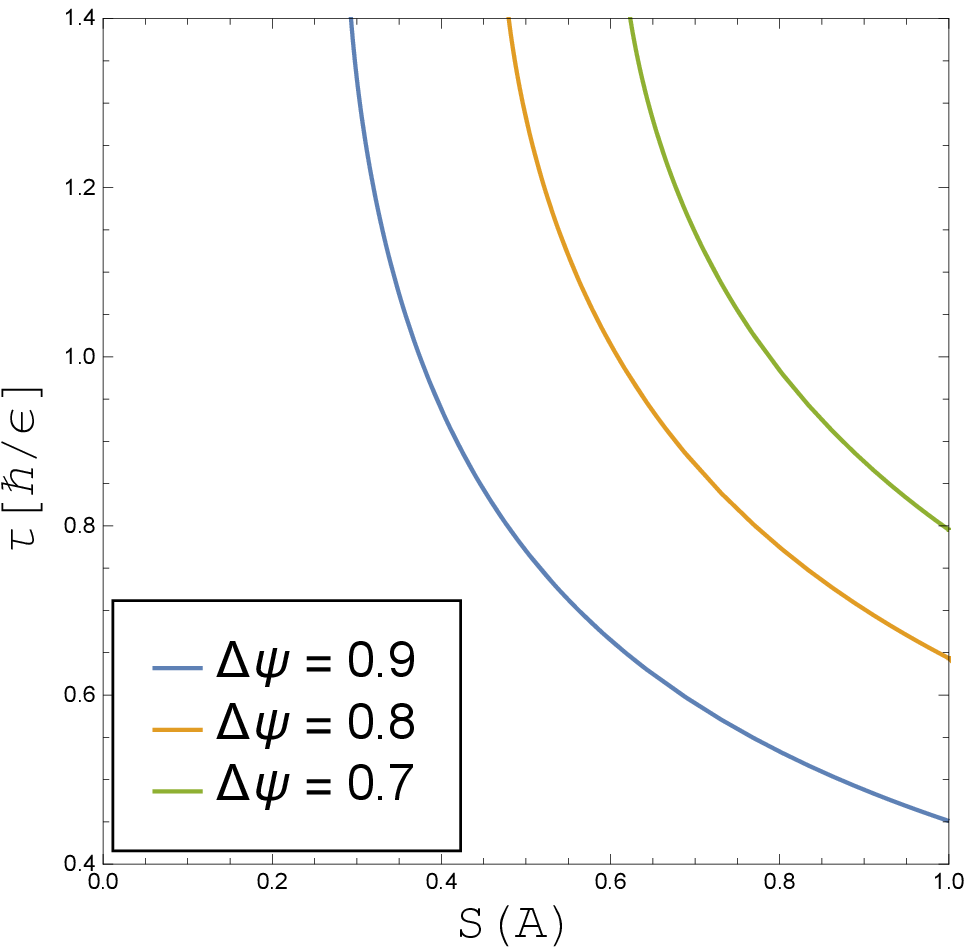}\hspace{1cm}
    \includegraphics[scale=0.8]{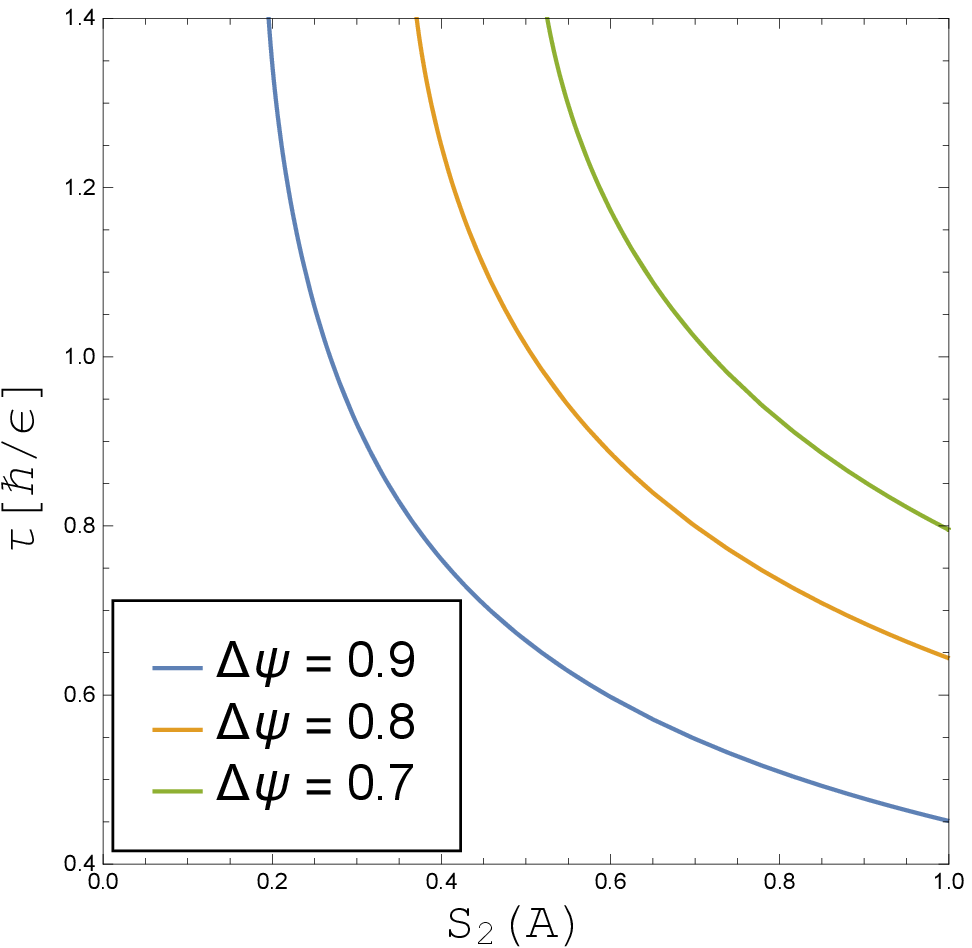}
    \caption{\textit{Upper panel:} Evolution time $\tau$ (in units of $\hbar/\epsilon$) as a function of entanglement entropy $S(A)$ for different distances $\Delta\psi$. \textit{Lower panel:} Evolution time $\tau$ (in units of $\hbar/\epsilon$) as a function of quadratic entanglement entropy $S_2(A)$ for different distances $\Delta\psi$.}
    \label{figure_speed}
\end{figure}

Using Eqs. \ref{S(A)} and \ref{tau}, we plot in the upper panel of Fig. \ref{figure_speed} the evolution time $\tau$ as a function of $S(A)$ for different values of $\Delta\psi$. Similarly, using Eqs. \ref{S2(A)} and \ref{tau}, we plot in the lower panel of Fig. \ref{figure_speed} the evolution time $\tau$ as a function of $S_2(A)$ for different values of $\Delta\psi$. From Fig. \ref{figure_speed}, we see that the more internal entanglement the system has, the further and faster its state vector can evolve.

\subsection{A single qubit clock}\label{Sec_Qubit clock}
To gain some initial intuition, we first consider the simplest case of a qubit clock that is entangled with the system as follows \cite{boette}:
\begin{equation}
    \ket{\Psi\rangle}=\frac{1}{\sqrt{2}}\left(\ket{0}_T\ket{\psi_0}_S+\ket{1}_T\ket{\psi_1}_S\right).
\end{equation}
The system evolves from the initial state $\ket{\psi_0}_S$ to the final state $\ket{\psi_1}_S$ while the qubit clock ticks from $\ket{0}_T$ to $\ket{1}_T$. The double ket notation of $\ket{\Psi}\rangle$ is just to remind us that it is the static ``global state" and not the evolving state of the system. The reduced density matrix of the system is
\begin{equation}
    \rho_S=\Tr_T(\ket{\Psi\rangle}\bra{\langle\Psi})=\frac{1}{2}(\ket{\psi_0}\bra{\psi_0}+\ket{\psi_1}\bra{\psi_1}),
\end{equation}
whose the nonzero eigenvalues are
\begin{equation}\label{p_pm Qubit}
    p_\pm=\frac{1\pm|\braket{\psi_1|\psi_0}|}{2},
\end{equation}
which can also be obtained from Eq. \ref{p_pm continuous no-interact} below by substituting $N=2$. The time-system entanglement entropy is then
\begin{equation}\label{E(T,S)}
    E(T,S)=-\Tr(\rho_S\log_2\rho_S)=-\sum_{k=\pm}p_k\log_2p_k.
\end{equation}
The quadratic time-system entanglement entropy is
\begin{equation}\label{E2(T,S)}
    E_2(T,S)=2(1-\Tr(\rho_S^2))=4p_+p_-.
\end{equation}

\begin{figure}[h!]
    \centering
    \includegraphics[scale=0.8]{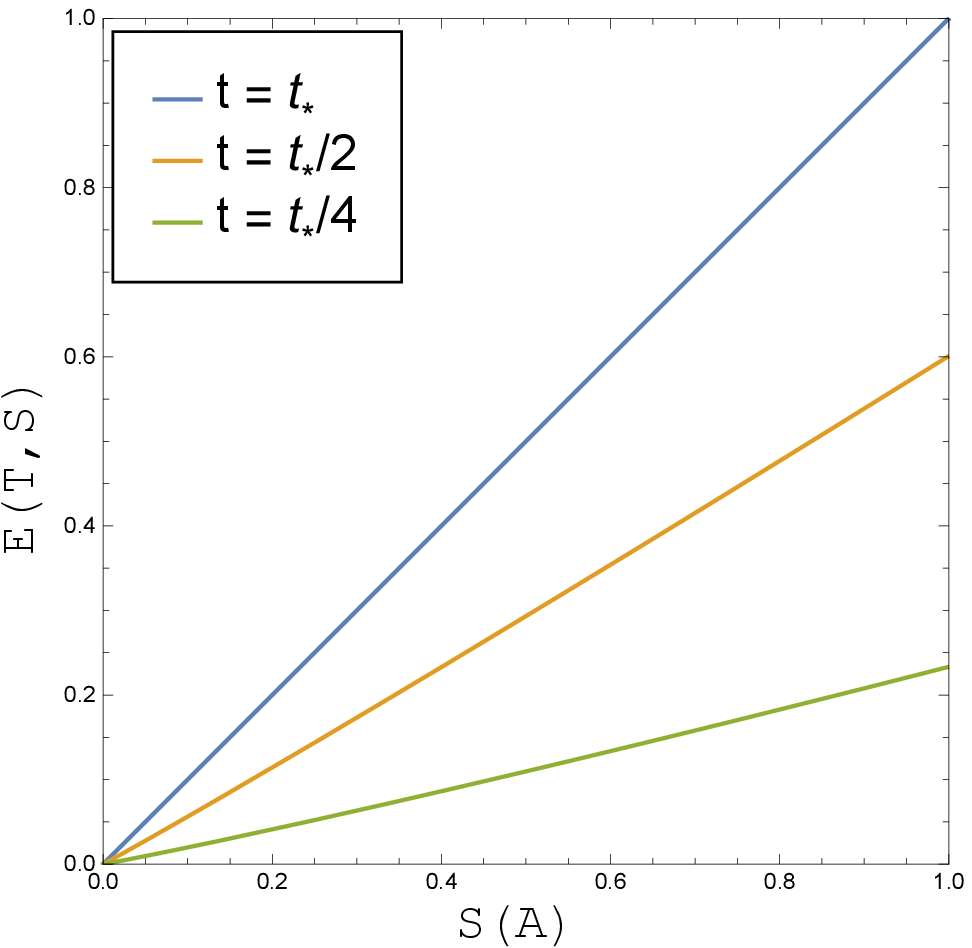}\hspace{1cm}
    \includegraphics[scale=0.8]{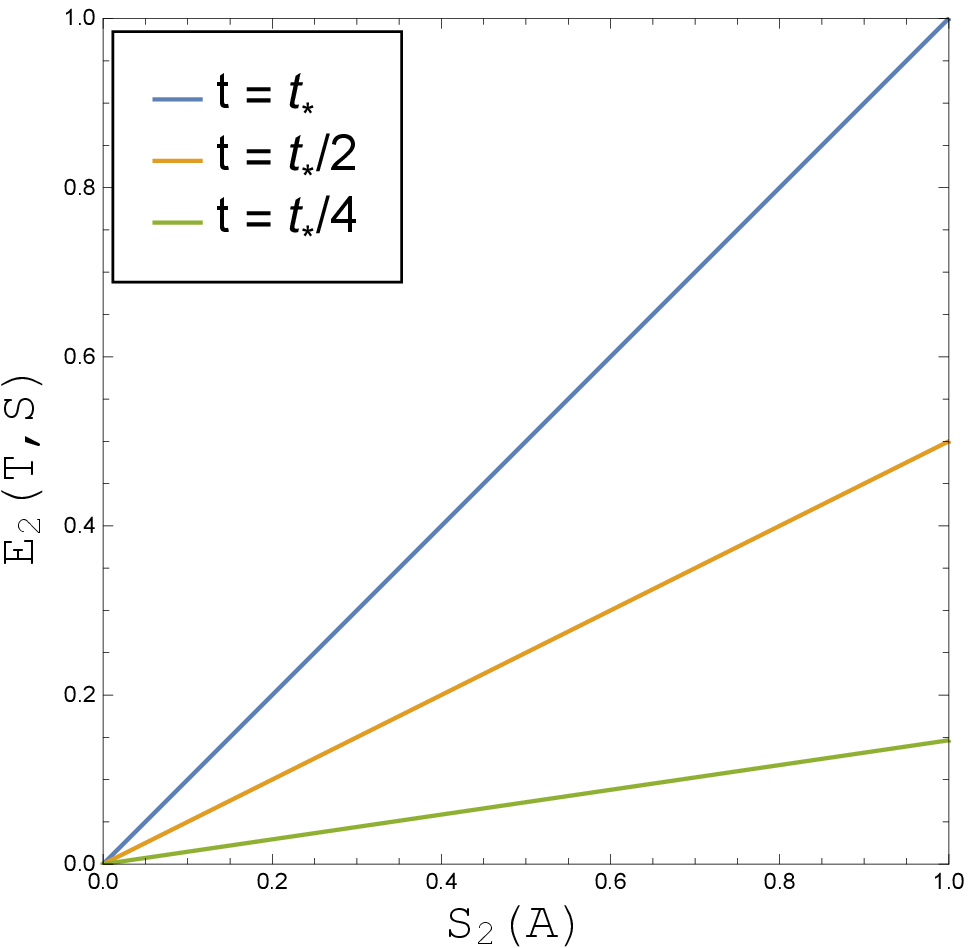}
    \caption{ \textbf{A single qubit clock}. \textit{Upper panel:} Time-System entanglement entropy, $E(T,S)$, as a function of internal entanglement entropy, $S(A)$, for different values of time. \textit{Lower panel:} Quadratic time-system entanglement entropy, $E_2(T,S)$, as a function of internal quadratic entanglement entropy, $S_2(A)$, for different values of time.}
    \label{figure_qubit}
\end{figure}

From Eq. \ref{p_pm Qubit}, we see that the more the system evolves, the more it becomes entangled with time. From Eq. \ref{fidelity}, we see that, for a given state with a fixed $|\alpha|^2$, the minimum time at which the fidelity is minimum is \cite{29}
\begin{equation}\label{t_* non-interacting}
    t_*^{NI}=\frac{\pi\hbar}{2\epsilon}.
\end{equation}
This time corresponds to the maximum time-system entanglement. The superscript ``NI" indicates the non-interacting case to distinguish it from the interacting case discussed later in Sec. \ref{sec_interacting}. Using Eqs. \ref{E(T,S)} and \ref{S(A)}, we can plot $E(T,S)$ as a function of $S(A)$. Similarly, using Eqs. \ref{E2(T,S)} and \ref{S2(A)}, we can plot $E_2(T,S)$ as a function of $S_2(A)$. These two plots are shown in Fig. \ref{figure_qubit}.

From Fig. \ref{figure_qubit}, we have the following remark. For a given time interval, we see that if the system has more entanglement within itself, then it becomes more entangled with time as it evolves. Combining this result with Fig. \ref{figure_speed}, we can now see the whole picture: the more internal entanglement the system has, the faster it can evolve. Therefore, during a fixed time interval, the state vector can travel through a larger distance in Hilbert space and thus becomes more entangled with time. Both $E(T,S)$ and $E_2(T,S)$ have these features, though they have slightly different scales.

\subsection{Continuous quantum time}
We now generalize our results to the continuous limit where there are infinitely many steps of evolution between the initial and final states. A useful procedure is to consider the global state \cite{boette}
\begin{equation}
    \ket{\Psi\rangle}=\frac{1}{\sqrt{N}}\sum_{t'=0}^{N-1}\ket{t'}_T\otimes\ket{\psi(t')}_S,
\end{equation}
and then take the limit $N\rightarrow \infty$ in the end. The normalization factor here indicates that each moment of time is equally likely to be occupied by the system's evolution. By taking the limit $N\rightarrow\infty$, it is also true to say that the system's evolution generates an infinite-dimensional Hilbert space of quantum time. The state vector at time $t$ is
\begin{equation}
    \ket{\psi(t)}=\alpha\ket{00}+\beta e^{-i\theta_t}\ket{11},
\end{equation}
where $\theta_t\equiv 2\epsilon t/\hbar$ for brevity. Note that $\ket{\psi(t')}=\alpha\ket{00}+\beta e^{-i\theta_t\frac{t'}{N-1}}\ket{11}$, so the state vector moves from the initial state $\ket{\psi(0)}$ to the target state $\ket{\psi(t)}$ in $N-1$ steps.

\begin{widetext}
The reduced density operator of the system is
\begin{equation}
\rho_S=\Tr_T(\ket{\Psi}\rangle\langle\bra{\Psi})=|\alpha|^2\ket{00}\bra{00}+\gamma(t)\ket{00}\bra{11}+\gamma(t)^*\ket{11}\bra{00}+|\beta|^2\ket{11}\bra{11},
\end{equation}
\end{widetext}
where 
\begin{equation}
    \gamma(t)\equiv\alpha\beta^*\braket{T_1|T_0},
\end{equation}
\begin{equation}
    \ket{T_0}\equiv\frac{1}{\sqrt{N}}\sum_{t'=0}^{N-1}\ket{t'},
\end{equation}
\begin{equation}
    \ket{T_1}\equiv\frac{1}{\sqrt{N}}\sum_{t'=0}^{N-1}e^{-i\theta_t\frac{t'}{N-1}}\ket{t'}.
\end{equation}
The nonzero eigenvalues of the reduced density matrix $\rho_S$ are
\begin{equation}\label{p_pm continuous no-interact}
    p_\pm=\frac{1\pm\sqrt{1-4(|\alpha|^2(1-|\alpha|^2)-|\gamma(t)|^2)}}{2},
\end{equation}
where 
\begin{equation}
\begin{aligned}
    |\gamma(t)|^2&=\frac{|\alpha|^2(1-|\alpha|^2)}{N^2}\Bigg|\sum_{t'=0}^{N-1}e^{i\theta_t\frac{t'}{(N-1)}}\Bigg|^2\\
    &=\frac{|\alpha|^2(1-|\alpha|^2)}{N^2}\frac{\cos\left(\frac{N\theta_t}{N-1}\right)-1}{\cos\left(\frac{\theta_t}{N-1}\right)-1}.
\end{aligned}
\end{equation}
In the continuous limit, this reduces to
\begin{equation}\label{gamma_continuous}
    \lim_{N\rightarrow\infty}|\gamma(t)|^2=2|\alpha|^2(1-|\alpha|^2)\frac{1-\cos\theta_t}{\theta_t^2}.
\end{equation}
The entanglement measures can then be calculated as usual by taking the continuous limit $N\rightarrow\infty$:
\begin{equation}\label{E(T,S)_Continuous}
    E(T,S)=\lim_{N\rightarrow\infty}\left(-\sum_{k=\pm}p_k\log_2p_k\right),
\end{equation}
\begin{equation}\label{E2(T,S)_Continuous}
    E_2(T,S)=\lim_{N\rightarrow\infty}\left(4p_+p_-\right).
\end{equation}

\begin{figure}[h!]
    \centering
    \includegraphics[scale=0.8]{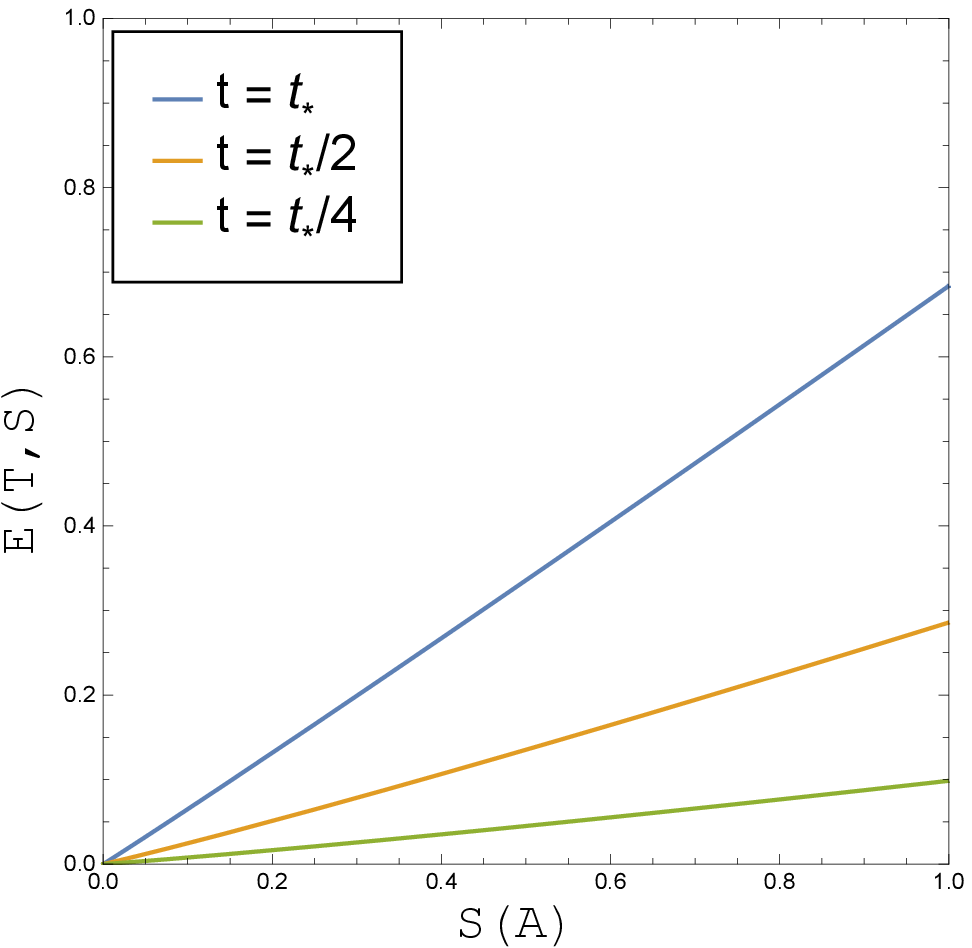}\hspace{1cm}
    \includegraphics[scale=0.8]{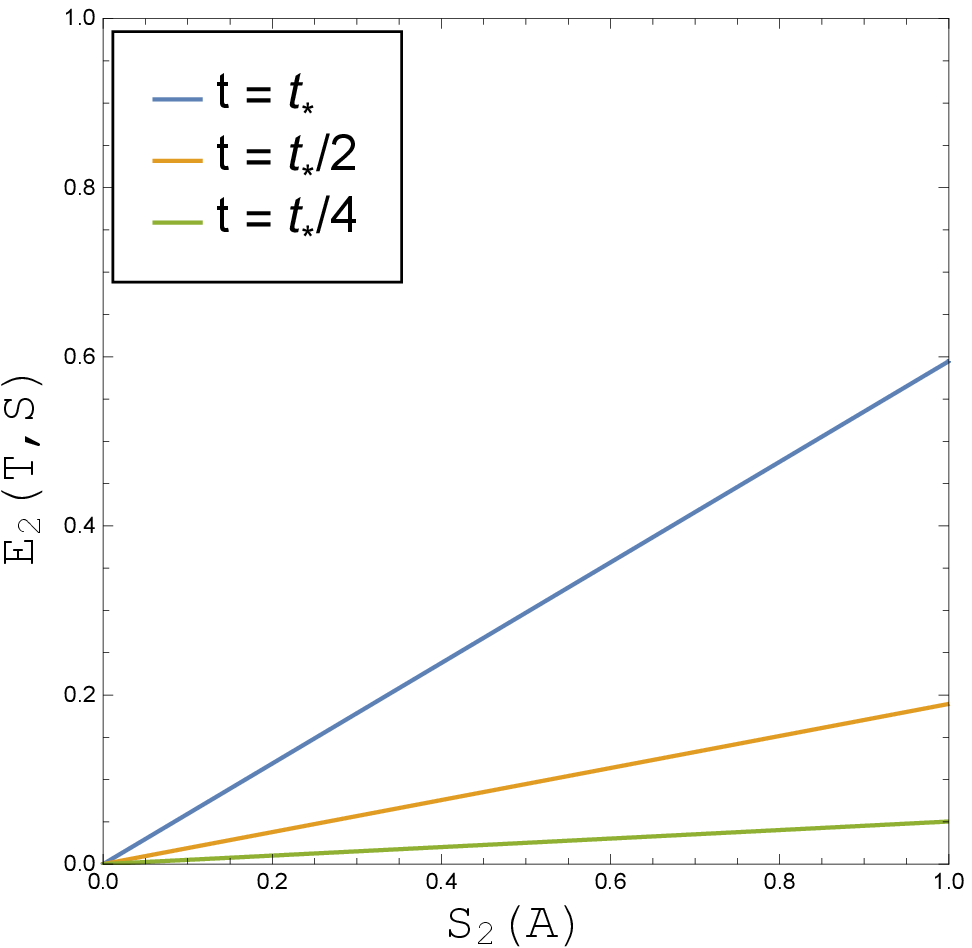}
    \caption{\textbf{Continuous quantum time.} \textit{Upper panel:} Time-System entanglement entropy, $E(T,S)$, as a function of internal entanglement entropy, $S(A)$, for different values of time. \textit{Lower panel:} Quadratic time-system entanglement entropy, $E_2(T,S)$, as a function of internal quadratic entanglement entropy, $S_2(A)$, for different values of time.}
    \label{figure_continuous}
\end{figure}

Using Eqs. \ref{E(T,S)_Continuous} and \ref{S(A)}, we can plot $E(T,S)$ as a function of $S(A)$. Similarly, using Eqs. \ref{E2(T,S)_Continuous} and \ref{S2(A)}, we can plot $E_2(T,S)$ as a function of $S_2(A)$. These plots are shown in Fig. \ref{figure_continuous}. From Fig. \ref{figure_continuous}, we have two remarks: (1) Similar to the single qubit clock case, increasing internal entanglement speeds up the evolution and makes the system more entangled with time; (2) In the continuous limit, we see that the time-system entanglement entropies are smaller than that of the single qubit clock case. In other words, if the state vector has to travel through more intermediate steps between the initial and final states, it becomes less entangled with time \cite{boette}.

Before moving on to the interacting case, some comments are in order. It is interesting to note that the elapsing time itself is not the factor to enhance time-system entanglement, but it is the ``distance of evolution". Alternatively, we can say that a faster-evolving state becomes more entangled with time as it evolves. A general state without internal entanglement can still evolve faster under the influence of some non-trivial Hamiltonian. However, as we mentioned in the introduction, quantum systems containing entangled subsystems are abundant in nature and are typical in quantum information protocols. A natural way to speed up the evolution of such systems is either increasing the internal entanglement within them for non-interacting systems (discussed above)  or increasing the strength of interaction for interacting systems (discussed below in Sec. \ref{sec_interacting}). Thus, the connection between the conventional internal entanglement within the system and the novel time-system entanglement is possible by using the speed of evolution as a mediator.

\section{Two interacting qubits}\label{sec_interacting}
We now consider the case where entanglement between two qubits is generated by interaction between them. We focus on computing the time-system entanglement entropy in the continuous limit. The total Hamiltonian is given by
\begin{equation}
    H_{total}=H_A\otimes I_B+I_A\otimes H_B+H_{int},
\end{equation}
where the local Hamiltonians $H_A$ and $H_B$ were defined in Eqs. \ref{local_Hamiltonian1} and \ref{local_Hamiltonian2} respectively. The interaction Hamiltonian is 
\begin{equation}
    H_{int}=-\lambda(\ket{1}\bra{0}\otimes\ket{1}\bra{0}+\ket{0}\bra{1}\otimes\ket{0}\bra{1}),
\end{equation}
where $\lambda>0$ is a coupling constant that has dimension of energy. This interaction Hamiltonian is capable of generating entanglement between two qubits. One can also add a constant term to the total Hamiltonian but that will only introduce an irrelevant overall phase factor. Consider the initial state to be a factorized state:
\begin{equation}
\ket{\psi(0)}=\ket{00}.   
\end{equation}
The state vector at time $t$ is given by the formal solution to the Schrodinger equation and is an entangled state:
\begin{equation}
\begin{aligned}
    \ket{\psi(t)}&=e^{-i\int H_{tot}dt/\hbar}\ket{00}\\
    &=\cos\frac{\lambda t}{\hbar}\ket{00}+i\sin\frac{\lambda t}{\hbar}\ket{11}\\
    &\equiv \cos\phi_t\ket{00}+i\sin\phi_t\ket{11} ,
\end{aligned}
\end{equation}
where we used the fact that $H_A\ket{0}_A=H_B\ket{0}_B=0$, $H_{int}\ket{00}=\ket{11}$ and $H_{int}\ket{11}=\ket{00}$. We also defined $\phi_t\equiv \lambda t/\hbar$ in the last line for brevity.

The global state is 
\begin{equation}
    \ket{\Psi}\rangle=\frac{1}{\sqrt{N}}\sum_{t'=0}^{N-1}\ket{t'}\otimes\ket{\psi(t')}.
\end{equation}
Here, $\ket{\psi(t')}=\cos\left(\phi_t\frac{t'}{N-1}\right)\ket{00}+i\sin\left(\phi_t\frac{t'}{N-1}\right)\ket{11}$, so the state vector moves from the initial state $\ket{\psi(0)}$ to the target state $\ket{\psi(t)}$ in $N-1$ steps. The reduced density operator of the system is
\begin{widetext}
\begin{equation}
\rho_S=\Tr_T(\ket{\Psi}\rangle\langle\bra{\Psi})=a(t)\ket{00}\bra{00}+c(t)\ket{00}\bra{11}+c^*(t)\ket{11}\bra{00}+b(t)\ket{11}\bra{11},
\end{equation}
where
\begin{equation}
    a(t)\equiv\frac{1}{N}\sum_{t'=0}^{N-1}\cos^2\left(\phi_t\frac{t'}{N-1}\right),
\end{equation}
\begin{equation}
    b(t)\equiv\frac{1}{N}\sum_{t'=0}^{N-1}\sin^2\left(\phi_t\frac{t'}{N-1}\right),
\end{equation}
\begin{equation}
    c(t)\equiv -\frac{i}{2N}\sum_{t'=0}^{N-1}\sin\left(2\phi_t\frac{t'}{N-1}\right).
\end{equation}
The nonzero eigenvalues of the system density operator $\rho_S$ are
\begin{equation}
    p_\pm=\pm\frac{1}{4N} \csc^2\left(\frac{\phi_t}{N-1}\right)\left[\pm N\left(1-\cos\left(\frac{2 \phi_t}{N-1}\right)\right)+2\sqrt{\sin^2\left(\frac{\phi_t}{N-1}\right)\sin^2\left(\frac{N\phi_t}{N-1}\right)}\right].
\end{equation}
\end{widetext}
In the continuous limit, this reduces to 
\begin{equation}\label{p_continuous}
    \lim_{N\rightarrow\infty}p_\pm=\frac{1}{2}\left(1\pm\frac{\sin\phi_t}{\phi_t}\right).
\end{equation}
Time-system entanglement entropy is then
\begin{equation}\label{E(T,S) interacting}
    E(T,S)=\lim_{N\rightarrow\infty}\left(-\sum_{k=\pm}p_k\log_2p_k\right).
\end{equation}

The fidelity between the initial and final states is
\begin{equation}
    \Delta\psi=|\braket{\psi(t)|\psi(0)}|=\cos\phi_t,
\end{equation}
and thus
\begin{equation}\label{phi_t}
    \phi_t=\arccos\Delta\psi.
\end{equation}
The minimum time required for the initial state to evolve to the orthogornal final state is
\begin{equation}\label{t_* interacting}
    t_*^I=\frac{\pi\hbar}{2\lambda}.
\end{equation}
The superscript ``I" indicates the interacting case to distinguish it from the non-interacting case discussed above in Sec. \ref{Sec_Qubit clock}. Comparing Eq. \ref{t_* interacting} with Eq. \ref{t_* non-interacting}, we see that the speed of evolution of two interacting qubits is greater than that of two non-interacting qubits if the interaction is sufficiently strong compared to the energy scale of the individual subsystems: $\lambda>\epsilon$.

\begin{figure}[h!]
    \centering
    \includegraphics[scale=0.88]{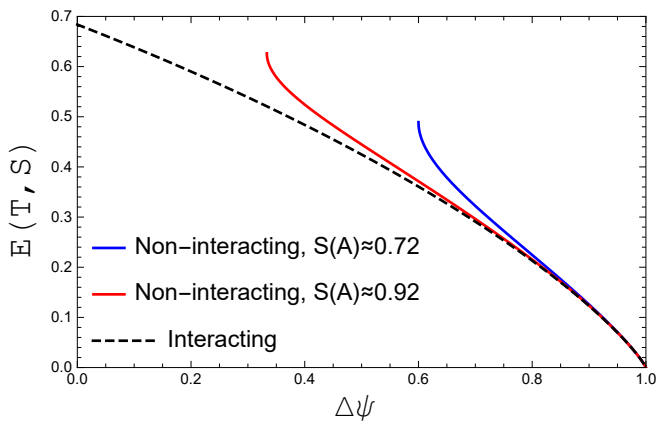}
    \caption{Time-system entanglement entropy E(T,S) as a function of fidelity $\Delta \psi$. \textit{Blue curve:} two non-interacting qubits with $S(A)\approx 0.72$ corresponding to $|\alpha|^2=1/5$ or $|\alpha|^2=4/5$. \textit{Red curve:} two non-interacting qubits with $S(A)\approx 0.92$ corresponding to $|\alpha|^2=1/3$ or $|\alpha|^2=2/3$. \textit{Dashed black curve:} two interacting qubits with time-dependent internal entanglement. The curve of two maximally entangled but non-interacting qubits ($S(A)=1$) coincides with the interacting curve, though it should be noted that the speed of evolution in the two cases is different (see Eqs. \ref{t_* non-interacting} and \ref{t_* interacting}).}
    \label{fig_interaction}
\end{figure}

Using Eqs. \ref{E(T,S) interacting} and \ref{phi_t}, we can plot $E(T,S)$ as a function of fidelity in Fig. \ref{fig_interaction}. The interacting case is represented by dashed black curve. From Fig. \ref{fig_interaction}, we see that $E(T,S)$ increases when fidelity decreases. That is because the more the system evolves, the more it becomes entangled with time. When the state vector reaches its maximum distance at $\Delta\psi=0$ (i.e. it evolves to an orthogonal state), the corresponding entropy it can acquire is $E_\perp (T,S)\approx 0.68$.

Also in Fig. \ref{fig_interaction}, we plot two solid colorful lines to represent the case of two non-interacting qubits by using Eqs. \ref{fidelity} and \ref{gamma_continuous}. These curves are cut at the maximum distances that the corresponding state vectors can travel. Among the non-interacting cases, we see that increasing internal entanglement speeds up the evolution, since the state vector can travel further in a fixed time interval and makes the system more entangled with time. If two non-interacting qubits are maximally entangled, its curve coincides with the interacting curve, though it should be noted that the speed of evolution in the two cases, in general, is different (see Eqs. \ref{t_* non-interacting} and \ref{t_* interacting}). If the interaction is sufficiently strong with $\lambda>\epsilon$ (i.e. $t_*^{I}<t_*^{NI}$), time-system entanglement entropy of the interacting system is always greater than that of the non-interacting system and we say that it can be entangled with time more effectively.

\section{Summary}\label{sec_conclusions}
Systems containing entangled subsystems are abundant in nature and are typical in quantum information protocols. At the fundamental level, the quantum nature of time is also a pressing and interesting topic. In this paper, we studied the entanglement between quantum time and a quantum system containing two entangled qubits. We considered the case of two initially entangled qubits evolving under local dynamics, as well as the case of two interacting qubits such that their internal entanglement is generated over time. In the first case, we obtained the main result that increasing internal entanglement speeds up the evolution and makes the system more entangled with time. For both cases, we showed the dependence of time-system entanglement entropy on the distance of evolution which is characterized by fidelity. We compared the two cases with each other and found that the interacting system can evolve faster than the non-interacting system if the interaction is sufficiently strong, and thus it is entangled with time more efficiently.

Besides the fundamental significance, our results could be useful to gain new insights of quantum time into black hole evaporation or cosmological perturbations in an expanding Universe, since we also have an evolving entangled bipartite system in those cases. For black hole, the total Hilbert space is decomposed on a spatial slice as $\mathcal{H}_{tot}=\mathcal{H}_{in}\otimes\mathcal{H}_{out}$, where $\mathcal{H}_{in}$ contains infalling degrees of freedom and $\mathcal{H}_{out}$ contains outgoing Hawking quanta. Interior and exterior degrees of freedom are entangled with each other \cite{Hawking}. For cosmological perturbations,  the Hilbert space can be decomposed in momentum space as $\mathcal{H}_{tot}=\mathcal{H}_{k>aH}\otimes\mathcal{H}_{k<aH}$, where $\mathcal{H}_{k>aH}$ contains subhorizon modes and $\mathcal{H}_{k<aH}$ contains superhorizon modes. Superhorizon modes are entangled with subhorizon modes \cite{loc2,EE}. In both cases, the system can be an evolving pure state that contains two entangled subsystems. It is therefore interesting to see how our idea can be applied to those cases and if quantum time can offer new insights into the information paradox in each case. We plan to investigate further along these lines.

\section*{Acknowledgements}
This special paper marks the 10-year anniversary of my professional career in physics. The amazing story started in August 2013 when I became an undergraduate in physics, although everyone thought that I would become a mathematician! Physics has given me everything I wanted: excitement of discovery of the unknowns, happiness of being successful,  interesting feelings of exploring new horizons beyond my home country, and a promising future ahead. I would like to take this opportunity to express my gratitude to my family, my teachers, my colleagues, and my friends for being an important part of this marvelous journey. It has been a long and proud journey to look back, but it is certainly not the end. I am super excited to see what the next decades will bring!


\begin{thebibliography}{99}
\bibitem{sean1} ChunJun Cao, Sean M. Carroll, and Spyridon Michalakis, \textit{Space from Hilbert space: Recovering geometry from bulk
entanglement}, Phys. Rev. D \textbf{95}, 024031 (2017).

\bibitem{sean2}  ChunJun Cao and Sean M. Carroll, \textit{Bulk entanglement
gravity without a boundary: Towards finding Einstein’s
equation in Hilbert space}, Phys. Rev. D \textbf{97}, 086003
(2018).

\bibitem{Giddings} S. B. Giddings, \textit{Quantum-First Gravity}, Found. Phys. \textbf{49}, 177–190 (2019).

\bibitem{aspect} Alain Aspect, Philippe Grangier, and Gérard Roger, \textit{Experimental Tests of Realistic Local Theories via Bell's Theorem}, Phys. Rev. Lett. \textbf{47}, 460 (1981).

\bibitem{joos} Erich Joos, H. Dieter Zeh, Claus Kiefer, Domenico Giulini, Joachim Kupsch, Ion-Olimpiu Stamatescu, \textit{Decoherence and the Appearance of a Classical World in Quantum Theory}, 2nd edition, Springer Berlin (2003).

\bibitem{Wigner1} Asher Peres, Petra F. Scudo, and Daniel R. Terno, \textit{Quantum Entropy and Special Relativity}, Phys. Rev. Lett. \textbf{88}, 230402 (2002).

\bibitem{Wigner2} Robert M. Gingrich and Christoph Adami, \textit{Quantum Entanglement of Moving Bodies}, Phys. Rev. Lett. \textbf{89}, 270402 (2002).

\bibitem{PaW} Don N. Page and William K. Wootters, \textit{Evolution without evolution: Dynamics described by stationary observables}, Phys. Rev. D \textbf{27}, 2885 (1983).

\bibitem{gio} Vittorio Giovannetti, Seth Lloyd, and Lorenzo Maccone,
\textit{Quantum time}, Phys. Rev. D \textbf{92}, 045033 (2015).

\bibitem{marletto} C. Marletto and V. Vedral, \textit{Evolution without evolution
and without ambiguities}, Phys. Rev. D \textbf{95}, 043510 (2017).


\bibitem{loc} Ngo Phuc Duc Loc, \textit{Time-system entanglement and special relativity}, Mod. Phys. Lett. A \textbf{39}, 2350183 (2024).


\bibitem{singh} Ashmeet Singh, \textit{Quantum Space, Quantum Time, and Relativistic Quantum Mechanics}, Quantum Stud.: Math. Found. \textbf{9}, 35-53 (2022).


\bibitem{diaz1}  N.L. Diaz, J.M. Matera, and R. Rossignoli, \textit{History state
formalism for scalar particles}, Phys. Rev. D \textbf{100}, 125020
(2019).

\bibitem{diaz2}  N.L. Diaz and R. Rossignoli, \textit{History state formalism for
Dirac’s theory}, Phys. Rev. D \textbf{99}, 045008 (2019).


\bibitem{diaz3} N.L. Diaz, J.M. Matera, and R. Rossignoli, \textit{Spacetime
quantum actions}, Phys. Rev. D \textbf{103}, 065011 (2021).


\bibitem{carlo}  Carlo Rovelli, \textit{Time in quantum gravity: An hypothesis},
Phys. Rev. D \textbf{43}, 442 (1991).


\bibitem{zeh}  H. D. Zeh, \textit{Time in quantum gravity}, Phys. Lett. A \textbf{126},
5 (1988).


\bibitem{sal}  H. Salecker and E. P. Wigner, \textit{Quantum Limitations of
the Measurement of Space-Time Distances}, Phys. Rev. \textbf{109}, 571 (1958).


\bibitem{kiefer} Claus Kiefer and Patrick Peter, \textit{Time in quantum cosmology}, Universe \textbf{8}, 36 (2022).


\bibitem{sebas}  Sebastian Gemsheim and Jan M. Rost, \textit{Emergence of
Time from Quantum Interaction with the Environment}, Phys. Rev. Lett. \textbf{131}, 140202 (2023).


\bibitem{exp1} Ekaterina Moreva, Giorgio Brida, Marco Gramegna, Vittorio Giovannetti, Lorenzo Maccone, and Marco Genovese, \textit{Time from quantum entanglement: An experimental illustration}, Phys. Rev. A \textbf{89}, 052122 (2014).

\bibitem{exp2}  Ekaterina Moreva, Marco Gramegna, Giorgio Brida,
Lorenzo Maccone, and Marco Genovese, \textit{Quantum time:
Experimental multitime correlations}, Phys. Rev. D \textbf{96},
102005 (2017).

\bibitem{exp3} Giulia Rubino, Lee A. Rozema, Francesco Massa, Mateus Ara´ujo, Magdalena Zych, Caslav Brukner, Philip Walther, \textit{Experimental entanglement of temporal order},
Quantum \textbf{6}, 621 (2022).



\bibitem{Hawking} Ahmed Almheiri, Thomas Hartman, Juan Maldacena, Edgar Shaghoulian, and Amirhossein 
 Tajdini, \textit{The entropy of Hawking radiation}, Rev. Mod. Phys. \textbf{93}, 035002 (2021).

\bibitem{loc2} Ngo Phuc Duc Loc, \textit{Unitary paradox of cosmological perturbations}, Int. J. Mod. Phys. D  \textbf{32},  2350050 (2023).

\bibitem{EE}  Suddhasattwa Brahma, Omar Alaryani, and Robert Brandenberger, \textit{Entanglement entropy of cosmological perturbations}, Phys. Rev. D \textbf{102}, 043529 (2020).

\bibitem{27} One can consider a more general entangled state with three independent coefficients. In which case, it was also shown that increasing internal entanglement speeds up the evolution \cite{batle}. We are here focusing on the Bell-type state to simplify the discussion and it is also more physically relevant for many practical purposes (e.g. see \cite{nelsen,susskind}).

\bibitem{boette} A. Boette, R. Rossignoli, N. Gigena, and M. Cerezo, \textit{System-time entanglement in a discrete-time model}, Phys. Rev. A \textbf{93}, 062127 (2016).


\bibitem{29} This minimum time is distinct from the so-called quantum speed limit (QSL)\cite{QSL1,QSL2}. The QSL is a bound on the minimum time required for a given state to evolve to an orthogonal state and it is obtained from the average and variance of energy. In our discussion, $t_*$ is the actual time of evolution for the initial state to evolve to the furthest possible state. It is minimum in the sense that it is obtained from $(2n+1)\pi\hbar/2\epsilon$ ($n\in\mathbb{Z}$) by setting $n=0$.


\bibitem{batle} J. Batle, M. Casas, A. Plastino, and A. R. Plastino, \textit{On the Connection Between Entanglement and the Speed of Quantum Evolution}, Phys. Rev. A \textbf{72}, 032337 (2005).

\bibitem{nelsen} Michael A. Nielsen and Isaac L. Chuang, \textit{Quantum Computation and Quantum Information}, Cambridge University Press (2010).

\bibitem{susskind} Juan Maldacena and Leonard Susskind, \textit{Cool horizons for entangled black holes}, Fortschr. Phys. \textbf{61}, No. 9, 781–811 (2013).

\bibitem{tanaka} T. Tanaka, G. Kimura, and H. Nakazato, \textit{Possibility of a minimal purity-measurement scheme critically depends on the parity of dimension of the quantum system}, Phys. Rev. A \textbf{87}, 012303 (2013).

\bibitem{QSL1} L. Mandelstam and I. G. Tamm, \textit{The Uncertainty Relation Between Energy and Time in Non-relativistic Quantum Mechanics}, J. Phys. (Moscow) \textbf{9}, 249 (1945).

\bibitem{QSL2} N. Margolus and L. B. Levitin, \textit{The maximum speed of dynamical evolution}, Physica D \textbf{120}, 188 (1998).

\end{thebibliography}
\end{document}